\documentclass[12pt]{article}
\usepackage{euscript}
\usepackage{amssymb}
\usepackage{amsmath}
\usepackage{graphicx}
\usepackage[small,sc,center]{titlesec}
\usepackage{indentfirst}

\newcommand{\nc}{\newcommand}
\nc{\ek}{E_\mathrm{K}}
\nc{\Frad}{F_\mathrm{rad}}
\nc{\fm}{\mbox{$.\!\!^{\mathrm m}$}}
\nc{\hp}{H_\mathrm{P}}
\nc{\K}{\,\mathrm{K}}
\nc{\Lrad}{L_\mathrm{rad}}
\nc{\mbol}{M_\mathrm{bol}}
\nc{\mzams}{M_\mathrm{ZAMS}}
\nc{\Req}{R_\mathrm{eq}}
\nc{\Teff}{T_\mathrm{eff}}
\nc{\Yc}{Y_\mathrm{c}}
\begin{document}

\begin{center}
\textbf{NONLINEAR PULSATIONS OF RED SUPERGIANTS}

\textbf{Yu. A. Fadeyev\footnote{E--mail: fadeyev@inasan.ru}}

\textit{Institute of Astronomy, Russian Academy of Sciences, Pyatnitskaya ul. 48, Moscow, 109017 Russia}

Received

\end{center}

\textbf{Abstract} --- Excitation of radial oscillations in population~I
($X=0.7$, $Z=0.02$) red supergiants is investigated using the solution
of the equations of radiation hydrodynamics and turbulent convection.
The core helium burning stars with masses $8M_\odot\le M\le 20M_\odot$ and
effective temperatures $\Teff < 4000$K are shown to be unstable against
radial pulsations in the fundamental mode.
The oscillation periods range between 45 and 1180 days.
The pulsational instability is due to the $\kappa$--mechanism in the hydrogen and
helium ionization zones.
Radial pulsations of stars with mass $M < 15M_\odot$ are strictly periodic with
the light amplitude $\Delta\mbol\le 0\fm5$.
The pulsation amplitude increases with increasing stellar mass and for
$M > 15M_\odot$ the maximum expansion velocity of outer layers is as high as
one third of the escape velocity.
The mean radii of outer Lagrangean mass zones increase due to nonlinear
oscillations by $\le 30$\% in comparison with the initial equilibrium.
The approximate method (with uncertainty of a factor of 1.5) to evaluate
the mass of the pulsating red supergiant with the known period of radial
oscillations is proposed.
The approximation of the pulsation constant $Q$ as a function of the
mass--to--radius ratio is given.
Masses of seven galactic red supergiants are evaluated using the
period--mean density relation.


\newpage
\section{INTRODUCTION}

Red supergiants are long--period variables with semiregular light variations on
a timescale $\gtrsim 10^2$ day.
The period--luminosity relation (Glass 1979; Feast et al. 1980) and
the linear theory of adiabatic oscillations (Stothers 1969, 1972) allow us
to suppose that such a type of variability is due to radial stellar pulsations.
At the same time together with semiregular variability some red supergiants exhibit
superimposed irregular light variations (Kiss et al. 2006).
The secondary stochastic variability is thought to be due to the large--scale
convection in the outer subphotospheric layers
(Stothers and Leung 1971; Schwarzschild 1975; Stothers 2010).
Red supergiants are also remarkable due to intensive mass loss revealed through
a large infrared excess indicating dust production in the stellar wind
(Verhoelst et al. 2009).

The period--luminosity relation of radially pulsating red supergiants is used for
determination of extra--galactic distances and in comparison with Cepheids
the red supergiants allow us to substantially extend the distance scale due to
their higher luminosities (Pierce et al. 2000; Jurcevic et al. 2000).
Application of the theory of stellar pulsation to the analysis of observed
variability of red supergiants allows us to verify some conclusions of
the stellar evolution theory in a way similar to that employed earlier
for Cepheids.
It should be noted also that the growing bulk of recent observations indicate that
the strong stellar wind of massive late--type supergiants is due to nonlinear
stellar oscillations (van Loon et al. 2008)

The nature of radial oscillations in red supergiants is still not completely clear yet.
The linear analysis of pulsational instability of red supergiants with
masses $15M_\odot\le M\le 30M_\odot$ was performed by 
Li and Gong (1994) and Guo and Li (2002).
According to their calculations radial oscillations of red supergiants
are due to instability of the fundamental mode and, perhaps,
the first overtone.
However the theoretical period--luminosity relation agrees only with
fundamental mode oscillations.

Nonlinear radial oscillations of red supergiants were considered only
in two studies.
In the first one (Heger et al. 1997) the authors investigated
radial oscillations of the red supergiant with mass $M = 15M_\odot$ at the
final stage of the core helium burning.
In the second work (Yoon and Cantiello 2010) the authors investigated
pulsational instability of red supergiants with masses
$15M_\odot\le M \le 40M_\odot$.
It should be noted that in both these studies
the self--exciting stellar oscillations were treated with modified methods
of stellar evolution calculation and effects of interaction between
pulsation motions and turbulent convection were not taken into account.

Below we present results of investigation of nonlinear pulsations of red supergiants
obtained from the self--consistent solution of the equations of radiation hydrodynamics
and turbulent convection.
The need for such an approach is due to the significant length and mass of
the outer convection zone involved in pulsation motions.
The treatment of convective heat transport uses the solution of the diffusion--type
equations for the enthalpy and the mean turbulent energy obtained by Kuhfu\ss (1986)
for spherically--symmetric gas flows from the Navier--Stokes equation.
Thus, the results presented below deal with modelling the semiregular variability
and the secondary stochastic variability is not considered because this problem
is beyond the approximation of spherical geometry.
We consider the stars with masses at the zero--age main sequence
$8M_\odot\le\mzams\le 20M_\odot$ and initial fractional mass abundunces of
hydrogen and elements heavier than helium $X=0.7$ and $Z=0.02$.

\section{INITIAL CONDITIONS}

The problem of self--exciting stellar oscillations is the Cauchy problem
for equations of hydrodynamics with initial conditions corresponding
to the hydrostatic and thermal equilibrium.
In the present study the initial conditions were taken from the evolutionary
models of stars at the core helium burning.
Methods of stellar evolution calculations are discussed in our previous papers
(Fadeyev 2011a, b).

\begin{figure}
\centerline{\includegraphics[width=15cm]{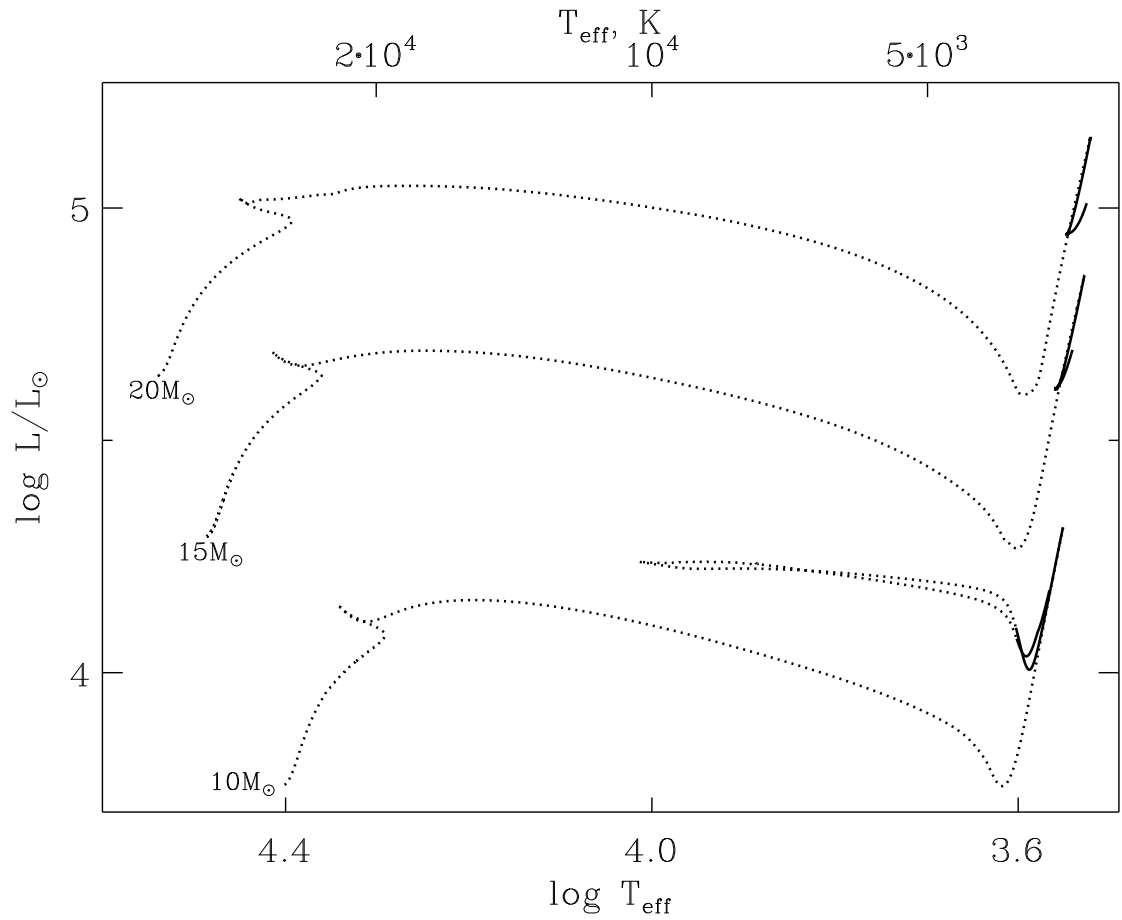}}
\caption{Evolutionary tracls of Population I stars with initial masses
$\mzams = 10$, 15 and $20M_\odot$ in the HR diagram.
Parts of tracks corresponding to the stage of the red supergiant
are shown in solid lines.}
\label{fig1}
\end{figure}

The evolutionary tracks in the Hertzsprung--Russel (HR) diagram of population~I
stars ($X=0.7$, $Z=0.02$) with initial masses $\mzams = 10$, 15 and $20M_\odot$
are shown in Fig.~\ref{fig1}.
Final points of tracks correspond to helium exhaustion in the stellar core
($Y_\mathrm{c}\approx 10^{-4}$).
Solid lines indicate parts of tracks when the star is the core helium burning
red supergiant with effective temperature $\Teff\le 4000\K$.

The star leaves the main sequence after core hydrogen exhaustion and
crosses the HR diagram to lower effective temperatures where becomes
the red supergiant with gradually increasing luminosity.
However, from the observational point of view the red supergiants with
gravitationally contracting helium core are not interesting since this
evolutionary stage proceeds in the Kelvin--Helmholtz time scale.
For example, for the star with initial mass $\mzams=10M_\odot$
the duration of gravitational contraction of the helium core is
$\approx 10^5$ years.

The luminosity ceases to increase when the triple alpha process becomes
the main energy source in the stellar center.
For the star witn initial mass $\mzams=10M_\odot$ the total duration of
thermonuclear helium burning is $\approx 2.8\cdot 10^6$ years and
in the beginning the star remains the red supergiant with luminosity
decreasing by a factor of two during $\approx 1.2\cdot 10^6$ years.
The star leaves the red supergiant domain when the central helium abundance
decreases below $\Yc\approx 0.52$ and its evolutionary track
loops the HR disgram to effective temperatures as high as $\Teff\approx 10^4$K.
The star becomes again the red supergiant when its central helium
abundance decreases to $\Yc\approx 0.05$ and the time of helium exhaustion
does not exceed $2\cdot 10^5$ years.
Therefore, of most interest is the initial stage of helium burning
during of which the luminosity of the red supergiant decreases.
Evolution of stars with initial masses $\mzams < 15M_\odot$
is nearly the same but proceeds in different time scales.

Evolutionary tracks of the core helium burning stars $\mzams\ge 15M_\odot$
do not loop in the HR diagram and all the time remain in the red supergiant domain.
For $\mzams=20M_\odot$ the core helium burning proceeds during $\approx 8.4\cdot 10^5$ years,
in the beginning during $\approx 3.6\cdot 10^5$ years the stellar luminosity decreasing
and then increasing to $L\approx 10^5L_\odot$.

All red supergiant evolutionary models used as initial conditions for
hydrodynamic computations are chemically homogeneous between the inner
boundary to the stellar surface.
Moreover,for the given value of $\mzams$ the abundances in the stellar envelope
do not change during the core helium burning.
This is due to the fact that the size of the outer convective zone is maximum
at the final stage of gravitational contraction of the helium core
just before ignition of the triple alpha process.
Depending on the mass and luminosity of the star
the radius of the inner boundary $r_0$ ranges within
$0.01 \lesssim r_0/\Req < 0.1$,
where $\Req$ is the radius of the upper boundary of the equilibrium model.

\section{RESULTS OF HYDRODYNAMIC COMPUTATIONS}

The method for the self--consistent solution of the equations of radiation hydrodynamics
and turbulent convection is described in our previous paper (Fadeyev, 2011b),
so that below we only discuss the results obtained.
Computations were carried out with the number of Lagrangean mass zones $500\le N\le 10^3$.
To be confident that the solution is independent of the inner boundary radius
$r_0$ and the number of Lagrangean zones $N$ some hydrodynamic models were
computed with several different values of these quantities.

Our hydrodynamic computations show that red supergiants with
initial masses $8M_\odot\le\mzams\le 20M_\odot$
are unstable against radial oscillations in the fundamental mode.
However depending on the value of $\mzams$ hydrodynamic models
demonstrate different behaviour both during the growth of instability
and after the limit amplitude attainment.
In particular, in stars with initial mass $\mzams < 15M_\odot$ 
the oscillation amplitude is always enough small, whereas for
$\mzams > 15M_\odot$ nonlinear effects play a substantial role.

The main properties of hydrodynamical models are summarized in Table~1,
where the bolometric luminosity $L$ and the effective temperature $\Teff$
correspond to the initial equilibrium model,
$\Yc$ is the fractional helium abundance in the stellar center,
$\Pi$ and $Q$ are the pulsation period and the pulsation constant in days,
$\eta = \Pi d\ln\ek/dt$ is the growth rate of the pulsation kinetic energy $\ek$.
Reciprocal of this quantity equals the number of pulsation periods
during which the kinetic energy increases by a factor of $e = 2.718\ldots$.
Ratios of the outer boundary mean radius $\langle R\rangle$ to the
equilibrium radius of the model $\Req$ are given in the last column
of Table~1 and show the role of nonlinear effects after the limit amplitude
attainment.

\begin{figure}
\centerline{\includegraphics[width=15cm]{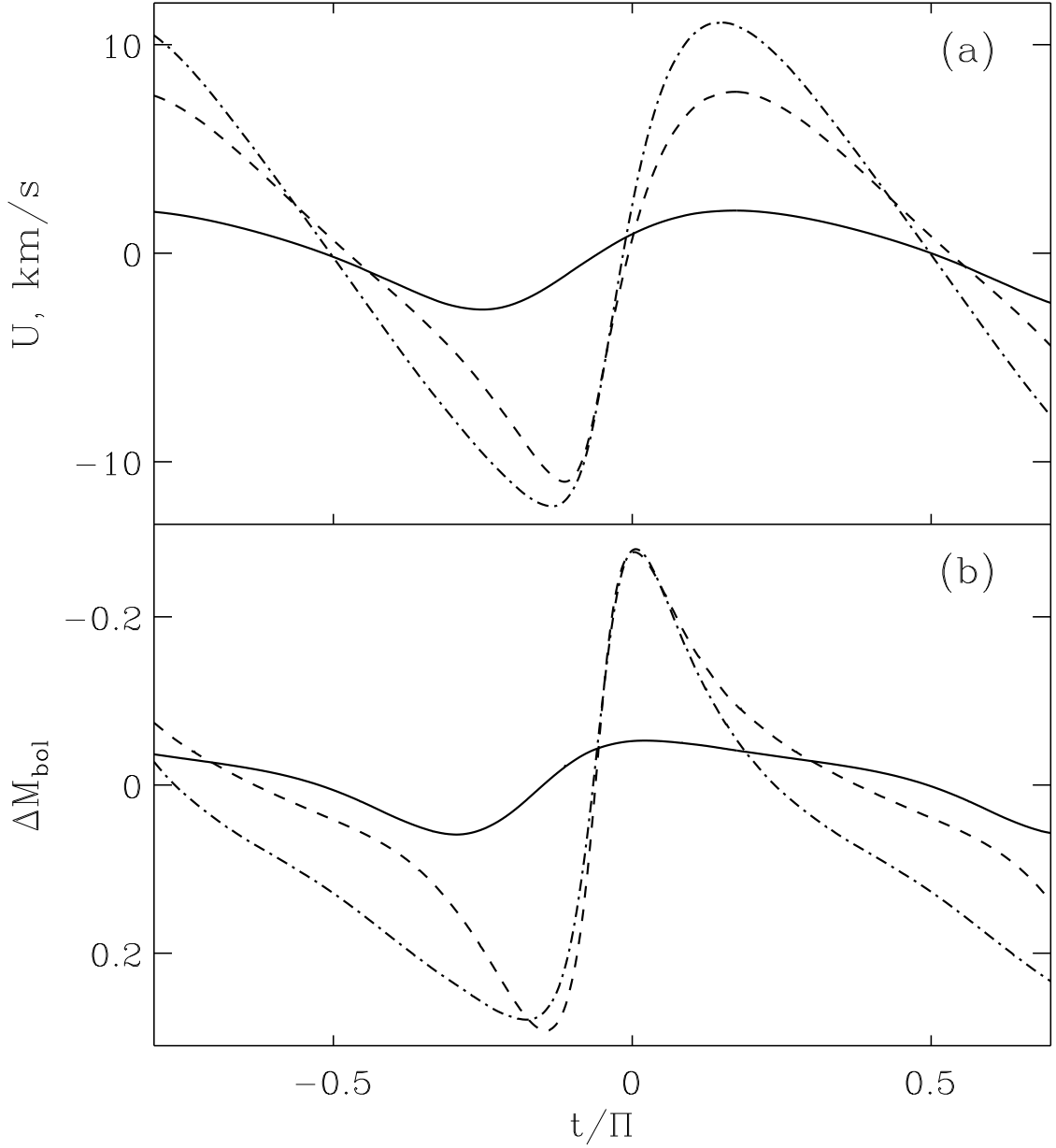}}
\caption{Variations of the gas flow velocity at the upper boundary $U$ (a)
and bolometric magnitude $\mbol$ (b) in red supergiants with initial mass
$\mzams = 10M_\odot$ and luminosity $L = 2\cdot 10^4$ (solid lines),
$1.5\cdot 10^4L_\odot$ (dashed lines) and $10^4L_\odot$ (dot--dashed lines).}
\label{fig2}
\end{figure}

The oscillation amplitude of red supergiants $\mzams=10M_\odot$
at the top of the evolutionary track is enough small and
the relative radial displacement of the upper boundary is $\Delta R/\Req=0.06$.
The oscillation amplitude gradually increases with decreasing luminosity
and in the point with minimum luminosity before the loop in the HR
diagram $\Delta R/\Req=0.13$.
Enhancement of the radial oscillation amplitude with decreasung luminosity
is illustrated in Fig.~\ref{fig2} where variations of the velocity at the
upper boundary $U$ and the bolometric magnitude $\mbol$ are shown for
$L=2\cdot 10^4$, $1.5\cdot 10^4$ and $10^4L_\odot$.

\begin{figure}
\centerline{\includegraphics[width=15cm]{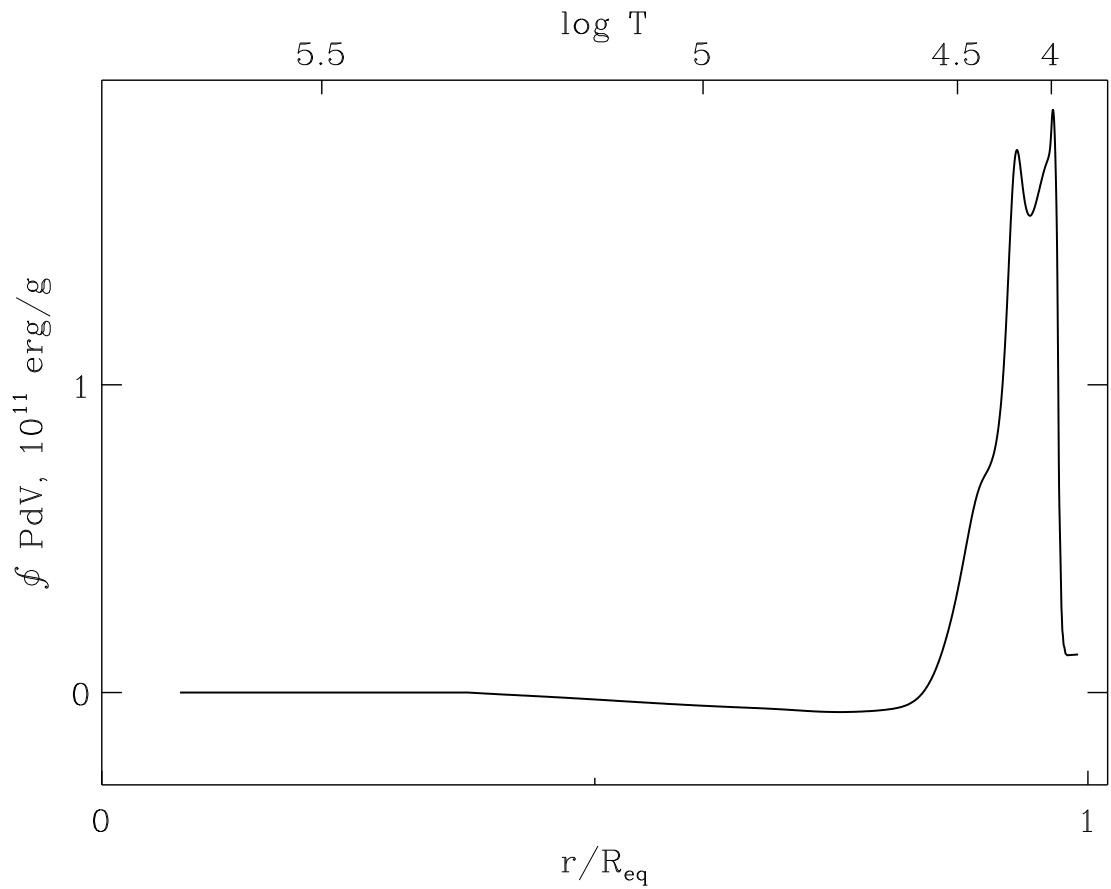}}
\caption{The radial dependence of the mechanical work done by a Lagrangean mass zone
over the pulsation period $\Pi$.}
\label{fig3}
\end{figure}

Fairly good repetition of small amplitude oscillations allows us
to calculate the mechanical work done by Lagrangean mass zones and
thereby to evaluate their contribution into excitation or damping
of instability.
The radial dependence of the mechanical work $\oint PdV$,
where $V$ is the specific volume and $P$ is the sum of gas, radiation
and turbulent pressure, is shown in Fig.~\ref{fig3} for the hydrodynamical
model $\mzams=10M_\odot$, $L=1.5\cdot 10^5L_\odot$.
The region of instability excitation ($\oint PdV > 0$) encompasses the layers
with temperature $1.2\cdot 10^4\lesssim T\lesssim 4\cdot 10^4$K
corresponding to the hydrogen and helium ionization zones.
In deeper layers ($T > 4\cdot 10^4$K) with fully ionized helium
the pulsational instability is damped ($\oint PdV < 0$).

To understand the physical mechanism of excitation of pulsational instability
let us consider variations of the gas density $\rho$, temperature $T$,
opacity $\kappa$ and luminosity $\Lrad = 4\pi r^2 \Frad$, where $\Frad$
is radiative flux, in Lagrangean mass zones of the hydrodynamical model.
In Fig.~\ref{fig4}(a) we give the plots of relative variations
$\delta\rho/\rho$, $\delta T/T$ and $\delta\kappa/\kappa$
in the layers of fully ionized helium
with temperature ranging within $4.9\cdot 10^4\K\le T\le 5.6\cdot 10^4\K$.
For the sake of convenience, the plots are arbitrarily shifted
along the vertical axis.
Coincidence of the maxima of density and temperature variations indicates
that oscillations are nearly adiabatic.
Decrease of opacity at maximum compression damps the instability
because, as seen in Fig.~\ref{fig4}(b), heat losses due to radiation
reach their maximum.

Variations of same quantities for the layer with temperature
$1.4\cdot 10^4\K\le T\le 1.7\cdot 10^4\K$ corresponding to
partial helium ionization are plotted in Fig.~\ref{fig5}.
Substantial phase shifts between maxima of density and temperature
indicate large nonadiabaticity of pulsation motions, whereas the delay
of the maximum temperature with respect to maximum compression
is the cause of the positive mechanical work.
Absorption of heat during compression is due to increase of opacity
and it is accompanied by decrease of the radiative flux.

Thus, damping of oscillations in the layers of fully ionized helium
and driving of pulsational instability in the hydrogen and helium ionization
zones are due to the $\kappa$--mechanism,
because effects of heat gains and losses are connected with
absorption and emission of radiation.
Low rates of the instability growth ($\eta\sim 10^{-2}$) and the small
limit cycle amplitude are due to the small fraction of radiation
in the total energy flux ($\Lrad\lesssim 10^{-2}L_r$).
Driving of pulsational instability at so small radiation fluxes
is due to the large amplitude of total luminosity variations.
As seen in Fig.~\ref{fig6}, the amplitude of luminosity variations
is largest in vicinity of the helium ionization zone.

Pulsational instability of red supergiants increases with increasing
initial stellar mass and for $\mzams > 15M_\odot$ nonlinear effects become
significant.
In Fig.~\ref{fig7} we give the temporal dependences of the kinetic energy
$\ek$ of the pulsating envelope and the radius of the upper boundary $R$
in units of the equilibrium radius $\Req$ for the red supergiant model
with initial mass $\mzams=16M_\odot$ and luminosity $L=8.5\cdot 10^4L_\odot$.
Compared to less massive supergiants this model demonstrates the growth
rate which is higher by an order of magnitude, whereas after the attainment
of the limiting amplitude the mean radii of outer
Lagrangean mass zones exceed their equilibrium values by nearly one third.
The amplitude ceases to grow in a transitional process encompassing roughly
a dozen of oscillation periods ($20 < t/\Pi < 30$).
During this time interval the sources of instability in the hydrogen and
helium ionization zones are balanced by damping of instability in layers
of fully ionized helium and shock radiative losses in the layers above
the photosphere.

\section{PERIOD--LUMINOSITY RELATION}

In the HR diagram red supergiants occupy the domain with relatively narrow
effective temperature range ($3000\K\lesssim\Teff\lesssim 4000\K$), so that
similar to Cepheids they exhibit correlation between the equilibrium luminosity
$L$ and the period of radial pulsations $\Pi$.
From observations such a correlation is known as the period--luminosity
relation.
The theoretical period--luminosity relation obtained in the present study is shown
in Fig.~\ref{fig8} for models of three evolutionary sequences with initial masses
$\mzams=10$, 15 and $20M_\odot$.
Hydrodynamical models of red giants at the evolutionary stage of decreasing
luminosity are shown by filled circles and the dashed lines show the linear fits
for each evolutionary sequence.

For stars with initial mass $\mzams=10M_\odot$ the evolutionary track loops the
HR diagram during helium burning and three models shown in Fig.~\ref{fig8} by
triangles correspond to the initial part of the loop with effective temperatures
$\Teff < 4000$K.
Therefore, one of the causes of the dispersion of points in the empirical
period--luminosity diagram is that some stars with masses $M < 15M_\odot$
leave the red supergiant domain and others return to it.

The chemical composition of outer layers involved in pulsation motions
does not change, whereas effects of mass loss for $M < 15M_\odot$ are
insignificant.
Therefore, red supergiants with almost exhausted helium in the stellar
center obey the same period--luminosity relation as stars in earlier phases
of helium burning.
This is illustrated in Fig.~\ref{fig8} where two models of stars $\mzams=10M_\odot$
with central helium abundances $\Yc=2.4\cdot 10^{-3}$ and $1.1\cdot 10^{-4}$
are shown by open circles.

As can be seen in Fig.~\ref{fig8}, the dispersion of the common correlation between
the period and luminosity of red supergiants is due to dependence of the both
luminosity and period on the stellar mass.
Therefore, one of the causes of dispersion on the empirical period--luminosity diagram
is due to different masses of observed stars.
It should also be noted that the mass--luminosity relation of red supergiants
and, therefore, the period--luminosity relation, depend on convective overshooting
and mass loss during the preceeding evolution.
An important role in the both mass--luminosity and period--luminosity relations
belongs also to abundances of heavy elements $Z$.
These effects, however, were beyond the scope of the present study.

\section{PERIOD--MASS DIAGRAM}

The equlibrium luminosity of the red supergiant during helium burning
changes within the ranges that depend on the initial stellar mass.
For example, in stars with $\mzams\approx 10M_\odot$ the luminosity decreases
by a half, whereas in red supergiants with initial mass $\mzams=20M_\odot$
the maximum to minimum luminosity ratio decreases to $\approx 1.6$.
The period of radial pulsations $\Pi$ changes simultaneosly with equilibrium
luminosity $L$.
Evolution of red supergiants $\mzams\le 12M_\odot$ between the
upper and lower luminosity limits is accompanied by the change of the
pulsation period by a factor of three.
The maximum to minimum period ratio decreases to a factor of two
for $\mzams=20M_\odot$ .

This property of red supergiants is illustarted in Fig.~\ref{fig9}
where for stars with initial masses $8M_\odot\le\mzams\le 20M_\odot$
we show the period--mass diagram.
Hydrodynamical models of stars at the top of the evolutionary track
are shown by filled circles.
Open circles indicate the red supergiant models with lower luminosity.
The diagram in Fig.~\ref{fig9} takes into account effects of mass loss and
along the vertical axis we give the mass values $M$ of evolving stars.
Evolution of the red supergiant corresponds to the displacement on the
diagram from right to left and then in the opposite direction.
For models $\mzams=10M_\odot$ and $20M_\odot$
this displacement is shown by arrows.

The period--mass diagram in Fig.~\ref{fig9} demonstrates the existence of the
limited area of mass and period values.
The borders of allowed masses and periods of radially pulsating
red supergiants can be approximately fitted as
\begin{equation}
\label{permas}
\log (M/M_\odot) = \left\{
\begin{array}{l}
0.153 + 0.365 \log\Pi \\
0.488 + 0.273 \log\Pi
\end{array}
\right.
\end{equation}
and in Fig.~\ref{fig9} they are shown by dashed lines.
Thus, from the observational estimate of the period $\Pi$
relations (\ref{permas}) allow us to evaluate the upper and lower mass limits
of the red supergiant.
For periods $\Pi \le 300$ the uncertainty of such an estimate is
about a factor of $\approx 1.5$.

\section{PULSATION CONSTANT}

The more exact value of the red supergiant mass can be obtained from
the period--mean density relation
\begin{equation}
\label{pmd}
\Pi = Q \left(R/R_\odot\right)^{3/2} \left(M/M_\odot\right)^{-1/2}
\end{equation}
provided that the pulsation period $\Pi$ and the mean stellar radius $R$
are known from observations.
The pulsation constant $Q$ is obtained from the theory of stellar pulsation
and in some cases can be expressed as a function of the stellar mass $M$ and
stellar radius $R$.
Substitution of this expression into the period--mean density relation (\ref{pmd})
allows us to obtain the explicit expression for the mass of the pulsating star.

The pulsation constants obtained in our hydrodynamical calculations of red supergiants
with initial masses $8M_\odot\le\mzams\le 20M_\odot$ and pulsation periods
$45~\mbox{сут}\le\Pi\le 1180~\mbox{сут}$
are shown in Fig.~\ref{fig10} as a function of mass--to--radius ratio
$f = (M/M_\odot)/(R/R_\odot)$.
The linear fit of the pulsation constant is written as
\begin{equation}
\label{pulq}
\log Q = -2.288 - 0.778 \log f
\end{equation}
and is shown in Fig.~\ref{fig10} by the dashed line.

Masses $M$ of seven galactic red supergiants evaluated from substitution
of (\ref{pulq}) into the period--mean density relation (\ref{pmd}) are given
in Table~2.
The periods $\Pi$ are taken from the General Catalogue of Variable Stars
(Samus et al. 2011).
The mean stellar radii were evaluated by Levesque et al. (2005) and
Josselin and Plez (2007).
In last two columns of Table~2 we give the lower $M_a$ and upper $M_b$
mass limits derived from (\ref{permas}).

Unfortunately, the existing estimates of mean radii of red supergiants are
still highly uncertain.
For example, the uncertainty in the effective temperature is $\approx 25$\%
(Josselin and Plez 2007), so that the uncertainty in the mean radius is
as high as $\approx 60$\%.
Therefore, the case when the stellar mass determined from the period--mean
density relation is beyond the interval $[M_a, M_b]$ should not be
considered as a contradiction.
For example, if we adopt that the radius of AD~Per is larger by 20\%
($R=548R_\odot$) then the stellar mass is $M=12.3M_\odot$,
that is within ranges given by (\ref{permas}).

\section{CONCLUSION}

Given in the previous section estimates of masses of seven
galactic red supergiants allow us to conclude that the theory of stellar
evolution is in an agreement with observational estimates of stellar radii.
To compare more stars with the theoretical period--luminosity relation
one should consider pulsational instability of red supergiants in the wider
interval of initial masses $\mzams$.

For more detailed theoretical period--luminosity relation one should consider
the role of some parameters used in evolutionary computations.
One of them is the overshooting parameter.
In the present study the evolutionary computations were done for the ratio
of the overshooting distance to the pressure scale height $l_\mathrm{ov}/\hp = 0.15$.
The need to know the role of this parameter is due to the dependence
of the mass--luminosity relation of helium burning stars on convective overshooting.

In stars with masses $M\ge 20M_\odot$ effects of mass loss during the
red supergiant evolutionary stage become significant.
In the present study the evolutionary calculations were done with mass loss rates
by Nieuwenhuijzen and de Jager (1990) however determination of the mass loss rate $\dot M$
as a function of fundamental stellar parameters remains disputable
(Mauron and Josselin, 2011)
Therefore, one should employ parametrization
of the expression for $\dot M$ and consider the mass--luminosity and
period--luminosity relations as a function of this parameter.

Another parameter which significantly affects the period--luminosity relation of
red supergiants is the mass fraction abundance of heavy elements $Z$.
Of special interest is the period--luminosity relation for $Z=0.008$
which is typical for the Large and Small Magellanic Clouds.

\newpage
\section*{REFERENCES}

\begin{enumerate}

\item Yu.A. Fadeyev, Pis'ma Astron. Zh. \textbf{37}, 13 (2011a) [Astron.Lett. \textbf{37}, 11 (2011a)].

\item Yu.A. Fadeyev, Pis'ma Astron. Zh. \textbf{37}, 440 (2011a) [Astron.Lett. \textbf{37}, 403 (2011b)].

\item M.W. Feast, R.M. Catchpole, D.S. Carter, et al., MNRAS \textbf{193}, 377 (1980).

\item I.S. Glass, MNRAS \textbf{186}, 317 (1979).

\item J.H. Guo and Y. Li, Astrophys.J \textbf{565}, 559 (2002).  

\item A. Heger, L. Jeannin, N. Langer, et al., Astron.Astrophys. \textbf{327}, 224 (1997).

\item E. Josselin and B. Plez, Astron.Astrophys. \textbf{469}, 671 (2007).

\item J.S. Jurcevic, M.J. Pierce, and G.H. Jacoby, MNRAS \textbf{313}, 868 (2000).

\item L.L. Kiss, Gy.M. Szab\'o, and T.R. Bedding, MNRAS \textbf{372}, 1721 (2006).

\item R. Kuhfu\ss, 1986, Astron.Astrophys. \textbf{160}, 116 (1986).

\item E. Levesque, P. Massey, K.A.G. Olsen, et al., Astrophys.J. \textbf{628}, 973  (2005).

\item Y. Li and Z.G. Gong, Astron.Astrophys. \textbf{289}, 449 (1994).

\item J.Th. van Loon, M. Cohen, J.M. Oliveira, et al., Astron.Astrophys. \textbf{487}, 1055 (2008).

\item N. Mauron and E. Josselin, Astron.Astrophys. \textbf{526}, in press (2011).

\item H. Nieuwenhuijzen and C. de Jager, Astron.Astrophys. \textbf{231}, 134 (1990).

\item M.J. Pierce, J.S. Jurcevic, and D. Crabtree, MNRAS \textbf{313}, 271 (2000).

\item N.N. Samus, E.V. Kazarovets, N.N. Kireeva, et al. General Catalogue of Variable Stars (2011).

\item M. Schwarzschild, Astrophys.J. \textbf{195}, 137 (1975).

\item R. Stothers, Astrophys.J \textbf{156}, 541 (1969).

\item R. Stothers, Astron.Astrophys. \textbf{18}, 325 (1972).

\item R. Stothers, Astrophys.J \textbf{725}, 1170 (2010).

\item R. Stothers, K.C. Leung, Astron.Astrophys. \textbf{10}, 290 (1971).

\item T. Verhoelst, N. Van der Zypen, S. Hony, et al., Astron.Astrophys. \textbf{498}, 127 (2009).

\item S.--C. Yoon and M. Cantiello, Astrophys.J. \textbf{717}, L62 (2010).

\end{enumerate}

\newpage
\begin{table}
\caption{Hydrodynamical models of red supergaints.}

\vskip 5pt
\begin{center}
\begin{tabular}{r|l|r|r|r|r|r|r|r}
\hline
 $\mzams/M_\odot$ & $M/M_\odot$ & $L/L_\odot$,& $\Teff$,& $\Yc$  & $\Pi$,&  $Q$,  & $\eta$ & $\langle R\rangle/\Req$ \\
                  &             &     $10^4$  &    K    &        &  day  &  day   &        &    \\
\hline
 20               &     19.53   &   14.029    &   3320  &  0.975 & 1178  & 0.1365 &        \\
                  &     18.36   &    9.380    &   3478  &  0.746 &  608  & 0.1058 &  0.27  &  1.24 \\      % 1400
                  &     17.47   &    8.782    &   3522  &  0.480 &  555  & 0.1028 &  0.30  &  1.11 \\[8pt] % 1530
 15               &     14.80   &    7.176    &   3370  &  0.968 &  651  & 0.1131 &  0.18  &  1.26 \\      % 2550
                  &     14.12   &    4.454    &   3556  &  0.680 &  310  & 0.0883 &  0.10  &  1.12 \\      % 2950
                  &     13.69   &    4.074    &   3620  &  0.379 &  262  & 0.0830 &  0.11  &  1.11 \\[8pt] % 3130
 10               &      9.88   &    2.055    &   3570  &  0.970 &  203  & 0.0867 &  0.01  &  1.15  \\     % 1730
                  &      9.76   &    1.507    &   3672  &  0.816 &  127  & 0.0747 &  0.03  &  1.10  \\     % 1900
                  &      9.66   &    1.102    &   3870  &  0.523 &   67  & 0.0615 &  0.03  &  1.04  \\[8pt]
  9               &      8.90   &    1.452    &   3610  &  0.973 &  154  & 0.0849 &  0.02  &  1.16  \\     % 2830
                  &      8.76   &    0.668    &   3958  &  0.580 &   45  & 0.0574 &  0.02  &  1.02  \\[8pt]% 3180
  8               &      7.02   &    0.962    &   3690  &  0.972 &  102  & 0.0765 &  0.02  &  1.13  \\     % 2650
\hline
\end{tabular}
\end{center}
\end{table}
\clearpage

\newpage
\begin{table}
\caption{Masses of galactic red supergiants.}

\vskip 5pt
\begin{center}
\begin{tabular}{r|l|r|r|r|r}
\hline
          & $\Pi$, day & $R/R_\odot$ & $M/M_\odot$ & $M_a/M_\odot$ & $M_b/M_\odot$ \\
\hline
 SU Per   & 533        & 780         & 17.1        & 14.1          & 17.1          \\
 W  Per   & 485        & 620         & 12.2        & 13.6          & 16.6          \\
 V602 Car & 635        & 860         & 17.7        & 15.0          & 17.9          \\
 AD Per   & 362.5      & 457         &  8.9        & 12.2          & 15.4          \\
 FZ Per   & 184        & 324         &  8.2        &  9.5          & 12.8          \\
 RW Cyg   & 550        & 676         & 12.9        & 14.2          & 17.2          \\
 SW Cep   &  70        & 234         &  9.8        &  6.7          &  9.8          \\
\hline
\end{tabular}
\end{center}
\end{table}
\clearpage

\newpage
\section*{Figure captions}

\begin{figure}
\centerline{\includegraphics[width=15cm]{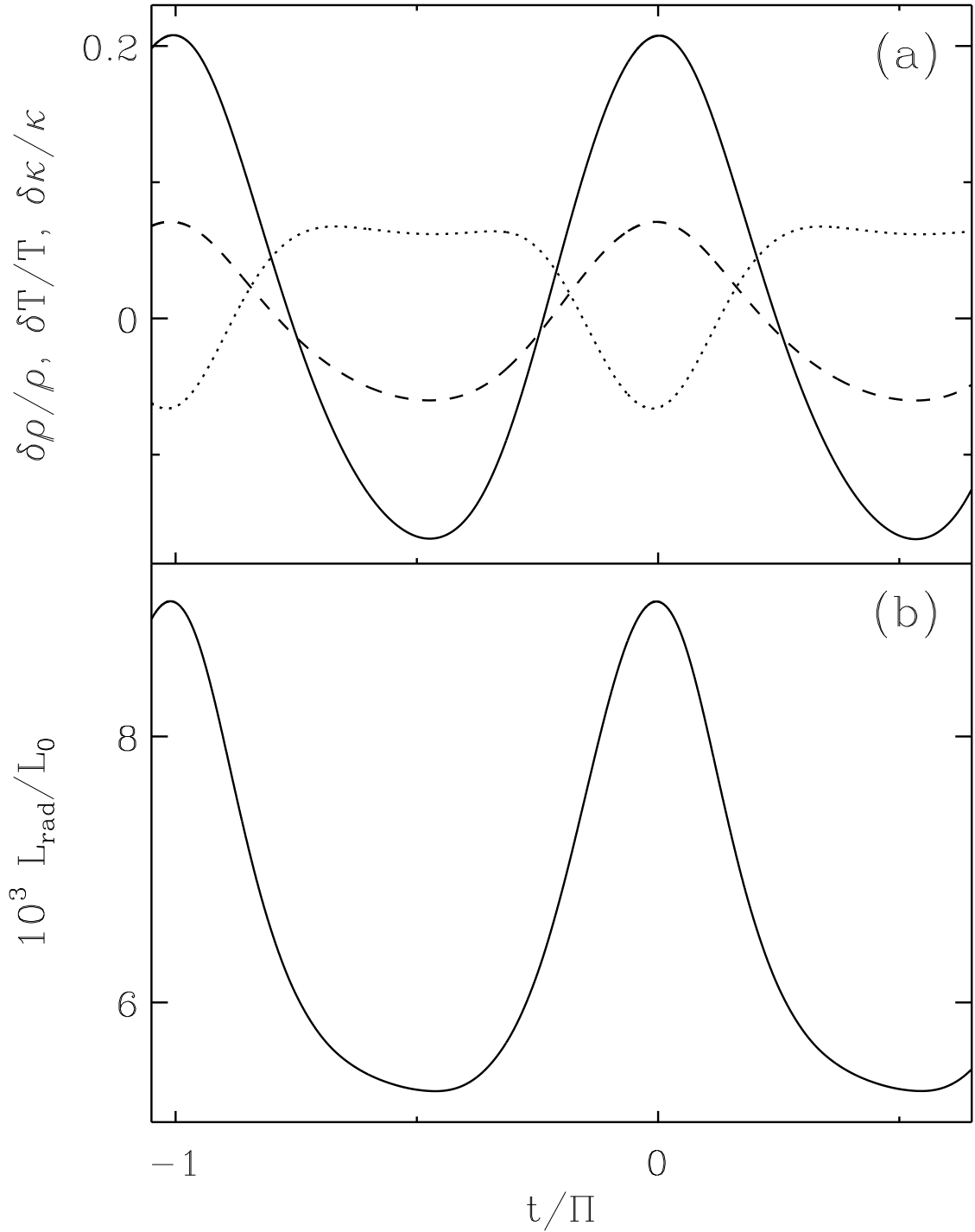}}
\caption{(а) -- Variations of gas density $\rho$ (solid line),
temperature $T$ (dashed line) and opacity $\kappa$ (dotted line)
in the layer of fully ionized helium;
(b) -- variations of radiative luminosity in units of the total equilibrium luminosity $L_0$.}
\label{fig4}
\end{figure}

\begin{figure}
\centerline{\includegraphics[width=15cm]{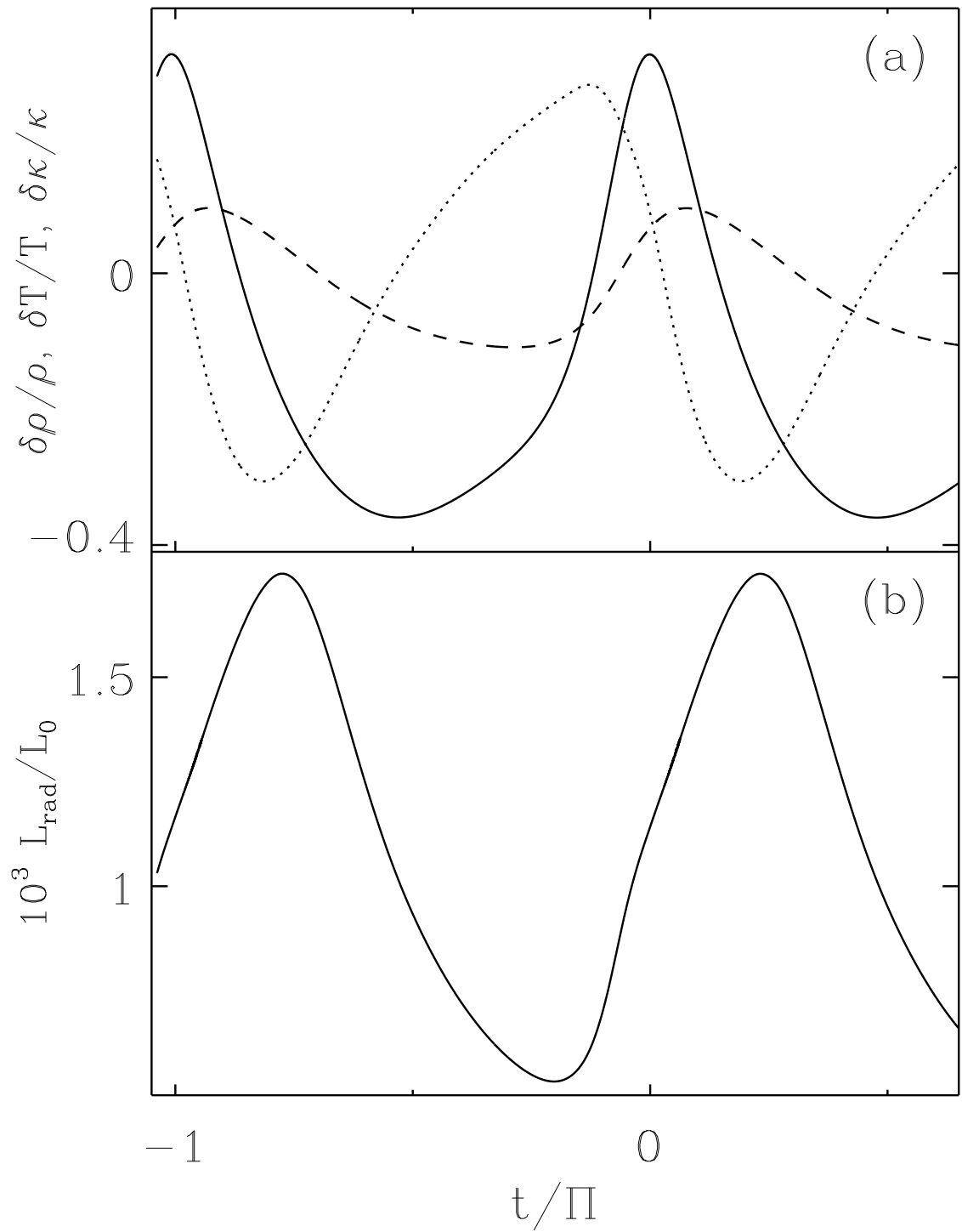}}
\caption{Same as Fig.~\ref{fig4} but for the layer with partial helium ionization.}
\label{fig5}
\end{figure}

\begin{figure}
\centerline{\includegraphics[width=15cm]{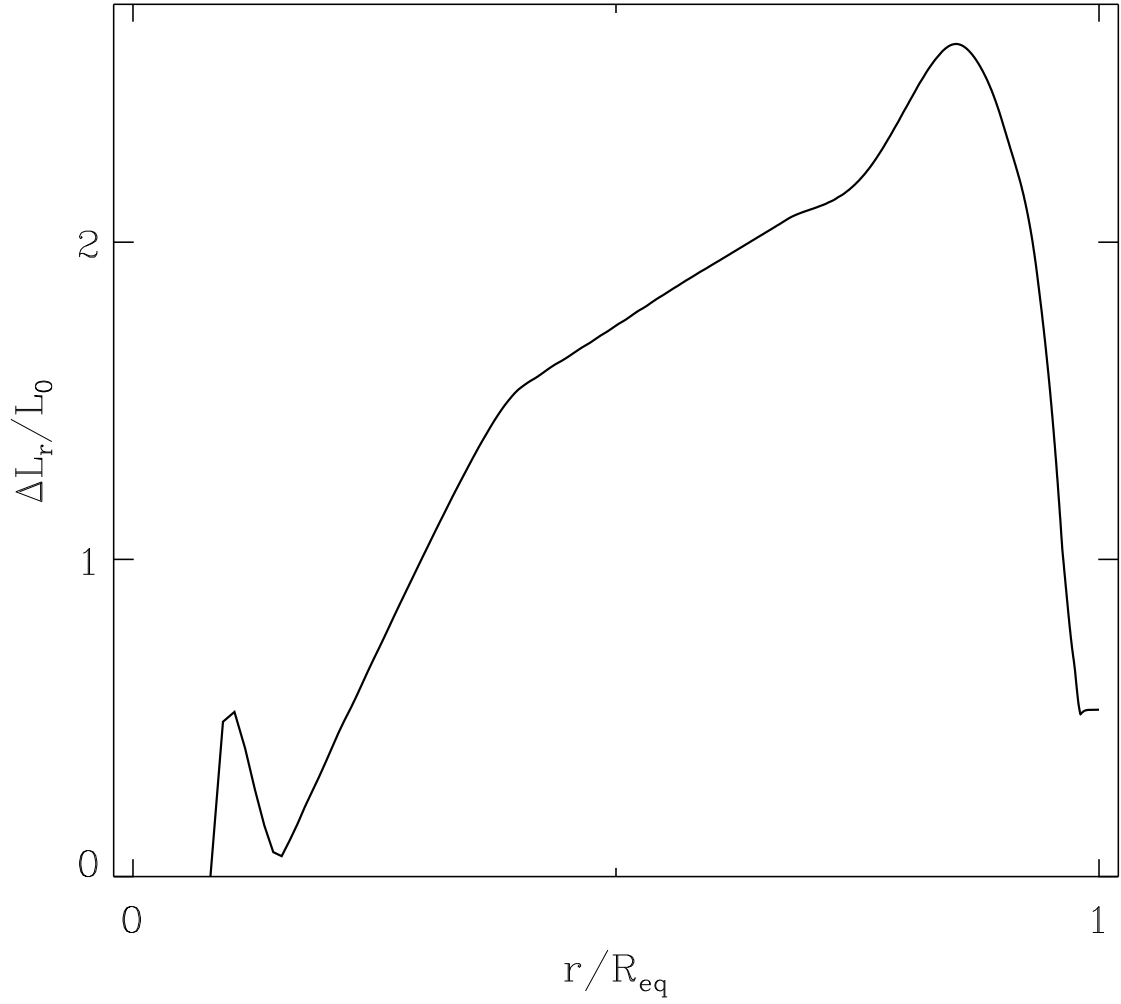}}
\caption{The amplitude of variations of the total luminosity $L_r$ expressed in
units of the equilibrium luminosity $L_0$ versus the radial distance from the stellar center.}
\label{fig6}
\end{figure}

\begin{figure}
\centerline{\includegraphics[width=14cm]{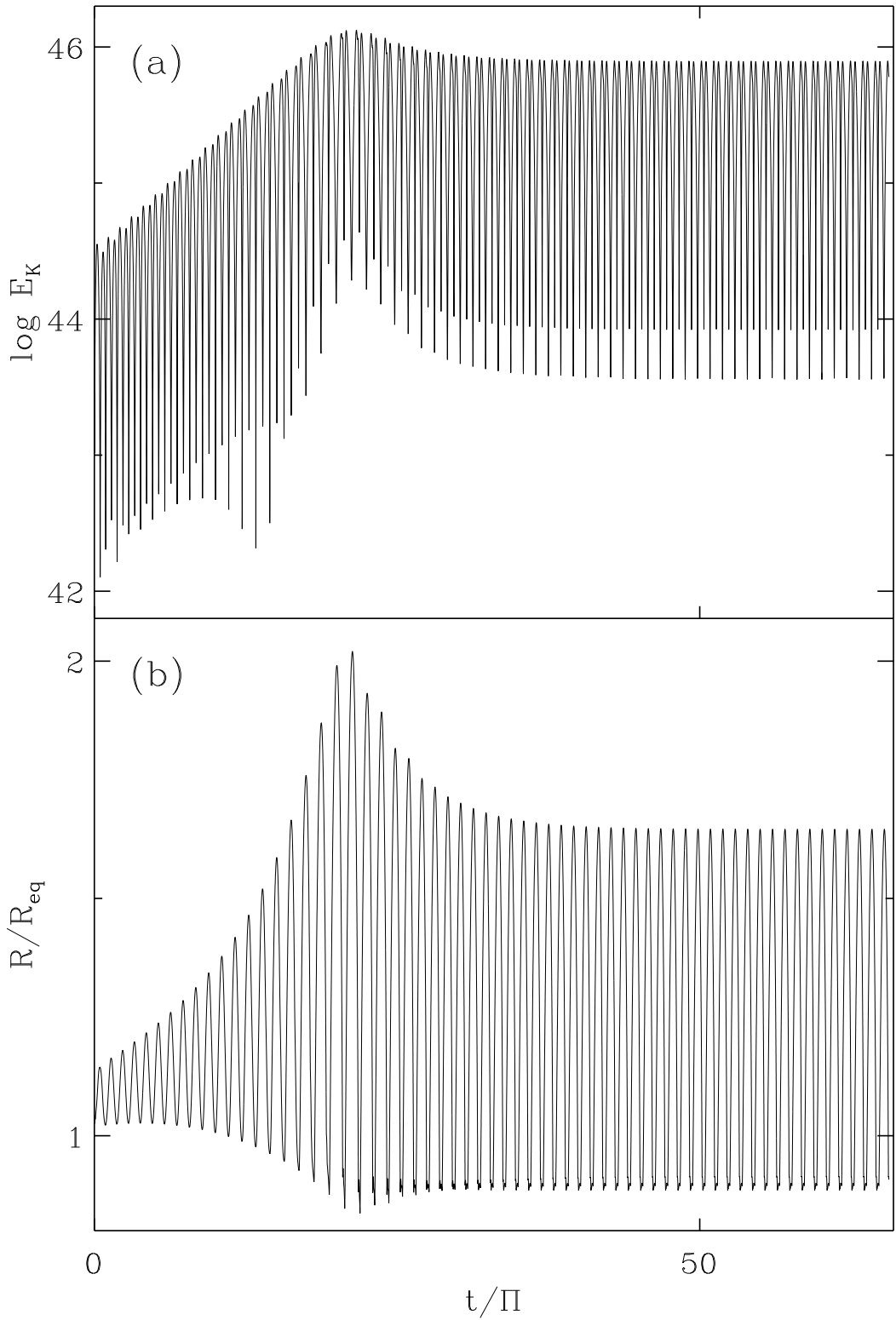}}
\caption{Kinetic energy of pulsation motions $\ek$ (a) and the upper boundary radius
$R$ (b) as a function of time $t$ for the hydrodynamical model $\mzams = 16M_\odot$,
$L=8.5\cdot 10^4L_\odot$.}
\label{fig7}
\end{figure}

\begin{figure}
\centerline{\includegraphics[width=15cm]{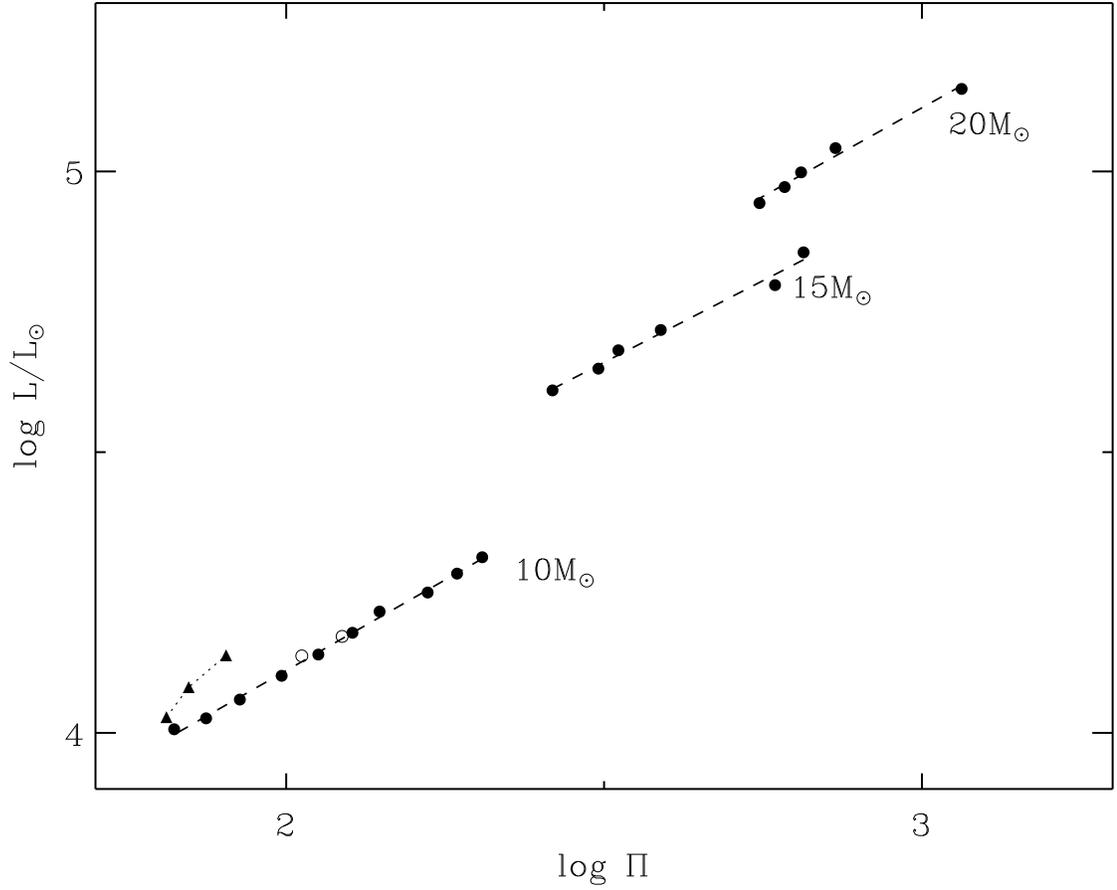}}
\caption{The period--luminosity diagram for red supergiants with initial masses
$\mzams=10$, 15 and $20M_\odot$.
Hydrodynamical models are represented by filled cicrles.
In triangles are shown hydrodynamical models of stars that leave the red supergiant domain.
In open circles are represented the hydrodynamical models of stars with central helium
abundance $\Yc\le 2.4\cdot 10^{-3}$.}
\label{fig8}
\end{figure}

\begin{figure}
\centerline{\includegraphics[width=15cm]{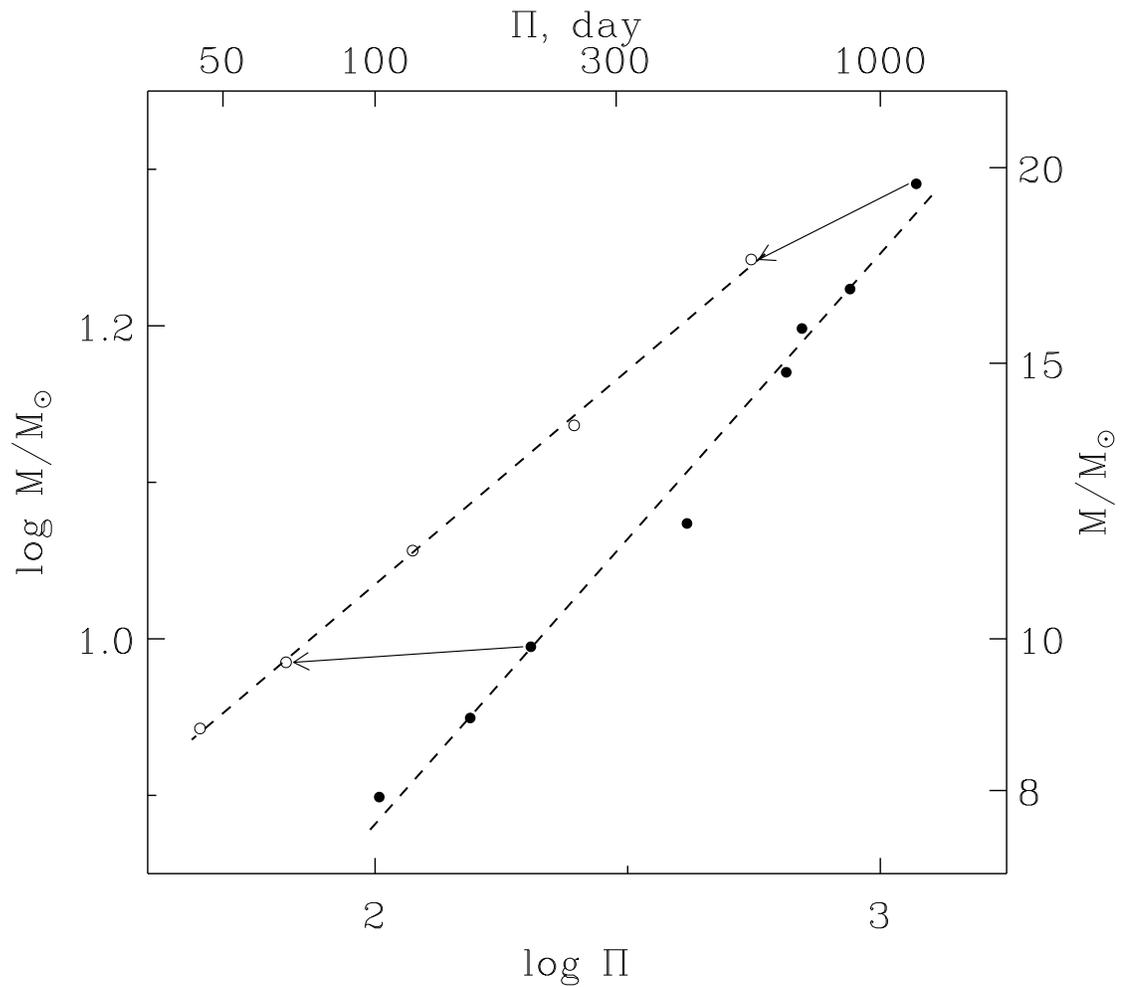}}
\caption{The period--mass diagram of red supergiants $8M_\odot\le\mzams\le 20M_\odot$.
Hydrodynamical models of stars with maximum and minimum equilibrium luminosity
are shown by filled and open circles, respectively.
Arrows indicate the direction of evolutionary change of the pulsation period $\Pi$
of stars $\mzams=10M_\odot$ and $20M_\odot$.
The region of allowed values of the radial pulsation periods is limited
by dashed lines.}
\label{fig9}
\end{figure}

\begin{figure}
\centerline{\includegraphics[width=15cm]{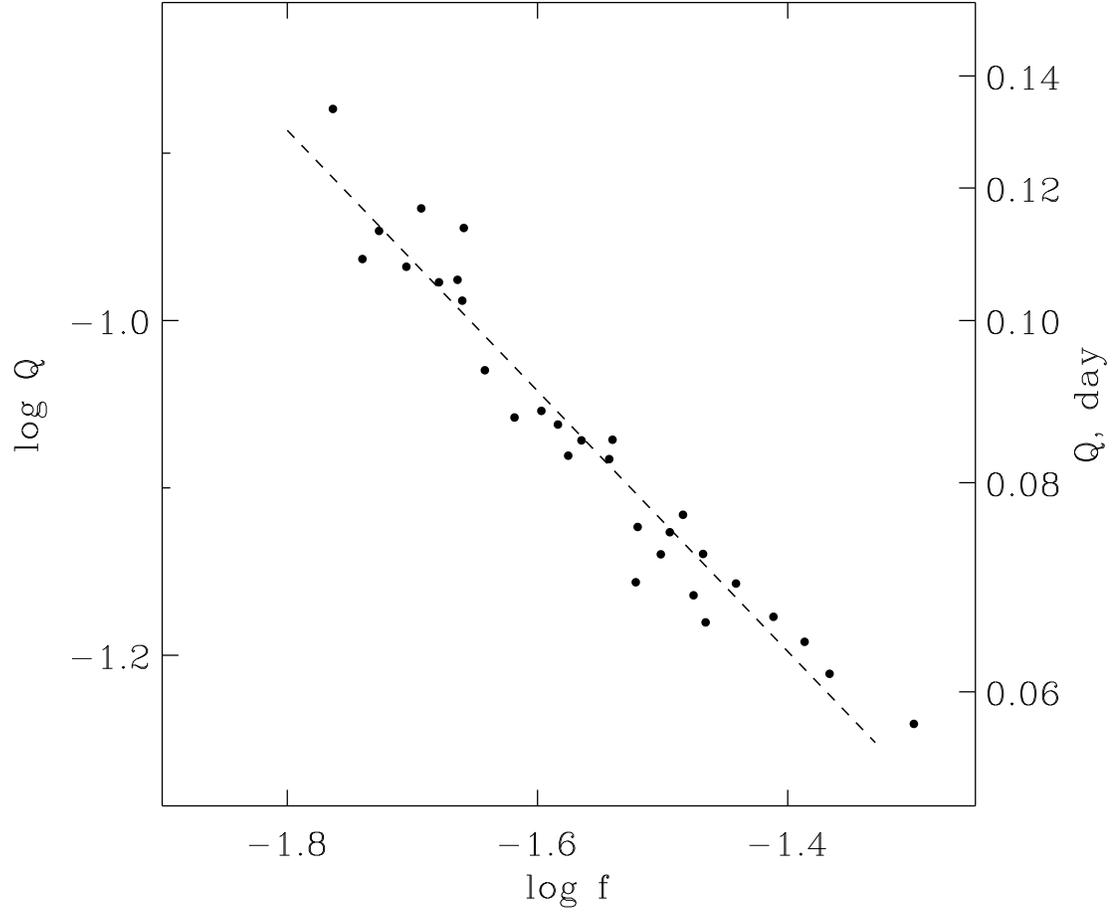}}
\caption{The pulsation constant $Q$ of red supergiants $8M_\odot\le\mzams\le 20M_\odot$
as a function of mass--to--radius ratio $f = (M/M_\odot)/(R/R_\odot)$.
Hydrodynamical models of red supergiants are shown by filled circles.}
\label{fig10}
\end{figure}

\begin{itemize}
\item[Fig. 1.]
Evolutionary tracls of Population I stars with initial masses
$\mzams = 10$, 15 and $20M_\odot$ in the HR diagram.
Parts of tracks corresponding to the stage of the red supergiant
are shown in solid lines.

\item[Fig. 2.]
Variations of the gas flow velocity at the upper boundary $U$ (a)
and bolometric magnitude $\mbol$ (b) in red supergiants with initial mass
$\mzams = 10M_\odot$ and luminosity $L = 2\cdot 10^4$ (solid lines),
$1.5\cdot 10^4L_\odot$ (dashed lines) and $10^4L_\odot$ (dot--dashed lines).

\item[Fig. 3.]
The radial dependence of the mechanical work done by a Lagrangean mass zone
over the pulsation period $\Pi$.

\item[Fig. 4.]
(а) -- Variations of gas density $\rho$ (solid line),
temperature $T$ (dashed line) and opacity $\kappa$ (dotted line)
in the layer of fully ionized helium;
(b) -- variations of radiative luminosity in units of the total equilibrium luminosity $L_0$.

\item[Fig. 5.]
Same as Fig.~\ref{fig4} but for the layer with partial helium ionization.

\item[Fig. 6.]
The amplitude of variations of the total luminosity $L_r$ expressed in
units of the equilibrium luminosity $L_0$ versus the radial distance from the stellar center.

\item[Fig. 7.]
Kinetic energy of pulsation motions $\ek$ (a) and the upper boundary radius
$R$ (b) as a function of time $t$ for the hydrodynamical model $\mzams = 16M_\odot$,
$L=8.5\cdot 10^4L_\odot$.

\item[Fig. 8.]
The period--luminosity diagram for red supergiants with initial masses
$\mzams=10$, 15 and $20M_\odot$.
Hydrodynamical models are represented by filled cicrles.
In triangles are shown hydrodynamical models of stars that leave the red supergiant domain.
In open circles are represented the hydrodynamical models of stars with central helium
abundance $\Yc\le 2.4\cdot 10^{-3}$.

\item[Fig. 9.]
The period--mass diagram of red supergiants $8M_\odot\le\mzams\le 20M_\odot$.
Hydrodynamical models of stars with maximum and minimum equilibrium luminosity
are shown by filled and open circles, respectively.
Arrows indicate the direction of evolutionary change of the pulsation period $\Pi$
of stars $\mzams=10M_\odot$ and $20M_\odot$.
The region of allowed values of the radial pulsation periods is limited
by dashed lines.

\item[Fig. 10.]
The pulsation constant $Q$ of red supergiants $8M_\odot\le\mzams\le 20M_\odot$
as a function of mass--to--radius ratio $f = (M/M_\odot)/(R/R_\odot)$.
Hydrodynamical models of red supergiants are shown by filled circles.

\end{itemize}

\newpage

\newpage

\newpage

\newpage

\newpage

\newpage

\newpage

\newpage

\newpage

\newpage

\end{document}